**SCARs: endogenous human stem cell-associated retroviruses and therapy-resistant malignant tumors**

Gennadi V. Glinsky[1]

[1] Institute of Engineering in Medicine

University of California, San Diego

9500 Gilman Dr. MC 0435

La Jolla, CA 92093-0435, USA

Correspondence: gglinskii@ucsd.edu

Web: http://iem.ucsd.edu/people/profiles/guennadi-v-glinskii.html

**Abstract**

Recent discoveries of endogenous human stem cell-associated retroviruses (SCARs) revealed consistent activation of specific endogenous retroviral elements in human preimplantation embryos and documented the essential role of the sustained retroviral activities in the maintenance of pluripotency, functional identity and integrity of naïve-state embryonic stem cells, and anti-viral resistance of the early-stage human embryos. SCARs activity have been implicated in seeding thousands' human-specific regulatory sequences in the hESC genome. Activation of specific SCARs, namely LTR7/HERVH and LTR5_Hs/HERVK, has been demonstrated in patients diagnosed with multiple types of cancer, autoimmune diseases, neurodegenerative disorders and it is likely associated with the emergence of clinically lethal therapy resistant death-from-cancer phenotypes in a sub-set of cancer patients diagnosed with different types of malignant tumors.

In human cells, retrotransposons' activity is believed to be suppressed to restrict the potentially harmful effects of mutations on functional genome integrity and to ensure the maintenance of genomic stability. Human embryonic stem cells (hESCs) and human embryos seem markedly different in this regard. In recent years, multiple reports demonstrate that retrotransposons' activity is markedly enhanced in hESC and human embryos and most active transposable elements (TEs) may be found among human endogenous retroviruses. Kunarso et al. [1] identified LTR7/HERVH as one of the most over-represented TEs seeding NANOG and POU5F1 binding sites throughout the human genome. Human endogenous retrovirus subfamily H (HERVH) RNA expression is markedly increased in hESCs [2, 3], and an enhanced rate of insertion of LTR7/HERVH sequences appears to be associated with binding sites for pluripotency core transcription factors [1, 2, 4] and long noncoding RNAs [5]. Expression of HERVH appears regulated by the pluripotency regulatory circuitry since 80% of LTRs of the 50 most highly expressed HERVH are occupied by pluripotency core transcription factors, including NANOG and POU5F1 [2]. Furthermore, TE-derived sequences, most notably LTR7/HERVH, LTR5_Hs/HERVK, and L1HS, harbor 99.8% of the candidate human-specific regulatory loci (HSRL) with putative transcription factor-binding sites (TFBS) in the genome of hESC [4].

The LTR7 subfamily is rapidly demethylated and upregulated in the blastocyst of human embryos and remains highly expressed in human ES cells [6]. In human ESC and induced pluripotent stem cells (iPSC), LTR7 sequences harboring the promoter for the downstream full-length HERVH-int elements, as well as LTR7B and LTR7Y sequences, were expressed at the highest levels and were the most statistically significantly up-regulated retrotransposons in human stem cells [7]. LTRs of human endogenous retrovirus subfamily H (HERVH), in particular, LTR7, function in hESC as enhancers and HERVH sequences encode nuclear non-coding RNAs, which are required for maintenance of pluripotency and identity of hESC [8]. Transient hyperactivation of HERVH is required for reprogramming of human cells toward induced pluripotent stem cells, maintenance of pluripotency and reestablishment of differentiation potential [9]. Failure to control the LTR7/HERVH activity leads to the differentiation-defective phenotype in neural lineage [9, 10]. The continuing activity of L1 retrotransposons may also contribute to the accelerated rate of creation of primate-specific transcription factor-binding sites during evolution because significant activities of both L1 transcription and transposition were recently reported in humans and other great apes [11].

Single-cell RNA sequencing of human preimplantation embryos and embryonic stem cells [12, 13] enabled identification of specific distinct populations of early human embryonic stem cells defined by marked activation of specific retroviral elements [14]. Consistent with definition of increased LTR7/HERVH expression as a hallmark of naive-like hESCs, a sub-population of hESCs and human induced pluripotent stem cells (hiPSCs) manifesting key properties of naive-like pluripotent stem cells can be genetically tagged and successfully isolated based on elevated transcription of LTR7/HERVH [15]. Targeted interference with HERVH activity and HERVH-derived transcripts severely compromises self-renewal functions [15]. Transactivation of LTR5_Hs/HERVK by pluripotency master transcription factor POU5F1 (OCT4) at hypomethylated long terminal repeat elements (LTRs), which represent the most recent genomic integration sites of HERVK retroviruses, induces HERVK expression during normal human embryogenesis, beginning with embryonic genome activation at the eight-cell stage, continuing through the stage of epiblast cells in preimplantation blastocysts, and ceasing during hESC derivation from blastocyst outgrowths [16]. Remarkably, Grow et al. [16] reported unequivocal experimental evidence demonstrating the presence of HERVK viral-like particles and Gag proteins in human blastocysts, consistent with the idea that endogenous human retroviruses are active and functional during early human embryonic development. Consistent with this idea, overexpression of HERVK virus-accessory protein Rec in pluripotent cells was sufficient to increase of the host protein IFITM1 level and inhibit viral infection [16], suggesting an anti-viral defense mechanism in human early-stage embryos triggered by HERVK activation.

Significantly, expression of HERVH-encoded long noncoding RNAs (lnc-RNAs) is required for maintenance of pluripotency and hESC identity [8]. In human ESC, 128 LTR7/HERVH loci with markedly increased transcription were identified [8]. It has been suggested that these genomic loci represent the most likely functional candidates from the LTR7/HERVH family playing critical regulatory roles in maintenance of pluripotency and transition to differentiation phenotypes in humans [4]. Conservation analysis of the 128 LTR7/HERVH loci with the most prominent expression in hESC demonstrates that none of them are present in Neanderthals genome, whereas 109 loci (85%) are shared with Chimpanzee [4]. Considering that Neanderthals' genomes are ~40,000 years old and Chimpanzee's genome is contemporary, these results are in agreement with the hypothesis that LTR7/HERVH viruses were integrated at these sites in genomes of primates' population very recently. Distinct patterns of expression of different sub-sets of transcripts selected from 128 LTR7/HERVH loci hyperactive in hESC are readily detectable in adult human tissues, including various regions of human brain [4]. These observations support the idea that sustained LTR7/HERVH activity may be relevant to physiological functions of human embryos and adult human organs, specifically human brain.

Taken together, these experiments conclusively established the essential role of the sustained, tightly controlled in the temporal-spatial fashion activity of specific endogenous retroviruses for pluripotency maintenance and functional identity of human pluripotent stem cells, including hESC and iPSC (**Figure 1**). Is this true for cancer stem cells as well and activation of human stem cell-associated retroviruses (SCARs) occurs in malignant tumors? Activation of the stemness genomic networks in human malignant tumors was linked with the emergence of clinically-lethal death-from cancer phenotypes in cancer patients, which are consistently associated with significantly increased likelihood of therapy failure, disease recurrence, and development of distant metastasis [17-26]. Gene expression signatures of the hESC genomic circuitry successfully identified therapy-resistant tumors in cancer patients diagnosed with multiple types of epithelial tumors [25, 26]. A cancer management guide has been developed to assist in clinical decision making process of therapy selection based on detection of "stemness" molecular and genetic markers in cancer cells [24]. Since these genotype/phenotype relationships between activation of stemness genomic networks and clinically-lethal therapy-resistant phenotypes of human cancers are readily detectable in the early-stage tumors from cancer patients diagnosed with multiple types of malignancies [17-26], it seems likely to expect that emergence of these tumors may be triggered by (or associated with) activation of endogenous human SCARs in cancer stem cells. One of the key molecular mechanisms of SCARs-mediated reprogramming of genomic regulatory networks is likely associated with functions of SCARs-derived long noncoding RNAs and human-specific TFBS (**Figure 1**). Consistent with this idea, experimental evidence revealing the mechanistic role in human cancer development one of the best studied to date LTR7/HERVH-derived long noncoding RNAs, namely linc-RNA-ROR [5, 27, 28], are beginning to emerge [29-33]. Comprehensive databases of transcriptionally active SCARs [8, 12-16], chimeric transcripts [14-16], and human-specific TFBS in the hESC genome [4] should greatly facilitate the follow-up mechanistic studies of human cancer.

Interestingly, expression of HERVK *Rec* has been demonstrated in the human placenta and various normal human fetal tissues [34], suggesting that activation of the Rec/IFITM1 anti-viral defense pathway may be important during the human pregnancy. Activation of HERVK (HML-2) expression has been reported in patients diagnosed with multiple sclerosis, rheumatoid arthritis, and various types of malignant tumors, including germ cell tumors, melanoma, lymphoma, breast, prostate, and ovarian cancers (reviewed in ref. 35). Expression of HERVH retroviruses has been reported in human cancer cell lines and clinical tumor samples [36-39]. Collectively, these findings underscore exciting new diagnostic and therapeutic opportunities for experimental and clinical development of molecularly-defined targeted interventions aiming at early detection and eradication of clinically-lethal sub-sets of malignant tumors in cancer patients.

**Figure legends**

**Figure 1.** Regulatory elements of pluripotency maintenance networks driven by sustained activity of endogenous human stem cell-associated retroviruses (SCARs). See text for further details and references. TFBS, transcription factor-binding sites; linc-RNA, long intergenic noncoding RNA; lnc-RNAs, long noncoding RNAs.

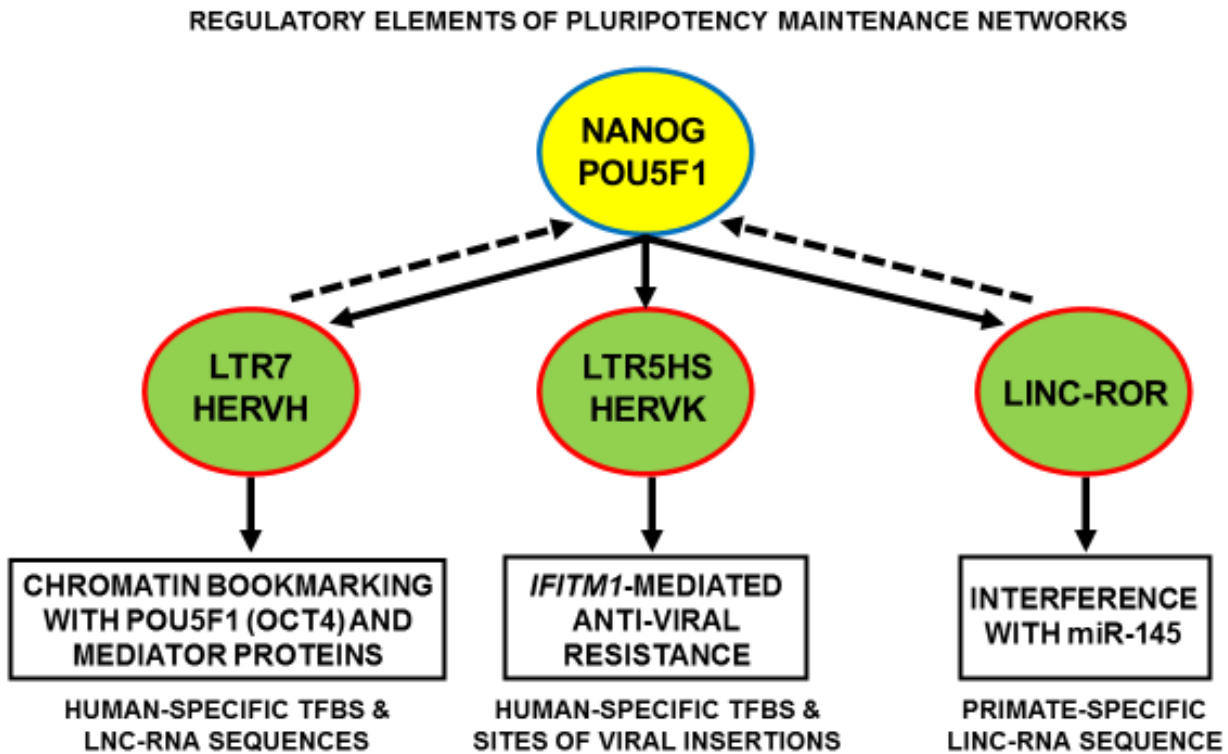